\documentclass[11pt]{article}
\usepackage{vietnam,epsfig}

\def\Journal#1#2#3#4{{#1} {\bf #2}, #3 (#4)}

\def\APJ{ApJ}
\def\AA{A\&A}

\def\be{\begin{equation}}
\def\ee{\end{equation}}
\def\bea{\begin{eqnarray}}
\def\eea{\end{eqnarray}}

\begin{document}
\vspace*{4cm}
\title{Preliminary results on galactic dark matter from \\the complete EROS2 microlensing survey}

\author{ P. TISSERAND  and  A. MILSZTAJN}

\address{
Dapnia, Service de Physique des Particules, CEA-Saclay, F-91191 Gif-sur-Yvette, France \\ 
for the EROS2 collaboration}

\maketitle\abstracts{
 The EROS-2 collaboration has conducted a survey of the Magellanic Clouds between July 1996
 and February 2003 (6.7 years), in order to search for microlensing phenomena due to putative
 machos (massive astronomical compact halo objects). About 55 million stars were monitored 
over 100 deg$^2$, typically with 400 to 500 measurements per star, simultaneously in two wide passbands.
 The full EROS-2 Magellanic Cloud data set was processed using the same program chain, and
 a photometric database was constructed. The lightcurves of the brightest 33 million stars were 
then searched for the characteristic signature of microlensing phenomena, in the wide timescale range
 of 1 to 1000 days (corresponding to macho masses ranging from about 0.01 solar mass to a few
 hundred solar masses). The sensitivity of this new analysis is better by a factor 4. The new EROS-2
 microlensing candidates towards the LMC and the corresponding (preliminary)
 final limits from EROS-2 on the macho content of the halo are presented; they are by far
 not numerous enough to account for the galactic dark matter. We have also followed, whenever
 possible, the microlensing candidates that were previously published by the EROS and MACHO
 collaborations. 1 MACHO and 2 EROS LMC candidates have shown subsequent variability and cannot be
 considered to be serious candidates.}


\section{Introduction}
Nowadays, the galactic dark matter problem is still not understood and has been confirmed from studies of many spiral galaxies.
 The rotation curves of their different components lead to a dynamical mass estimate 5-10 times higher that
 that of the visible components. Since 1986 and Paczy\'nski's proposal~\cite{pa}, gravitational
 microlensing has been proven to be a powerful probe of galactic halo dark matter under the form of machos.
 The signature of these objects is a transient amplification of the light from extragalactic stars by the gravitational
 lens effect when they are close to their line of sight. Events are extremely rare and millions of stars, towards the Large
 and Small Magellanic Clouds (LMC and SMC), have thus been monitored photometrically since 1990 by several
 experiments for a period of years in order to obtain a useful detection rate. Some microlensing candidates have
 been observed towards the two targets~\cite{lasserre2000,alcock2000a,afonso2003}: 9 by EROS (5 LMC, 4 SMC)
 and 18 by MACHO (16 LMC, 2 SMC), only 1 candidate is in common (97-SMC-1). As of 2003, both experiments have two different points of view:
 because of the unknown stellar variability background, EROS published an upper limit of the macho content of the halo,
 at 20\% (95\% C.L.) in the mass range $10^{-7}$ to 1 solar mass ($M_{\odot}$); and the MACHO collaboration suggest that their 
result can be understood as a signal of about 20\% of the halo mass with a mass for the lenses in the range
 0.15-0.9 $M_{\odot}$. Thus the two results were compatible. 

\section{The EROS experiment}
The EROS experiment (Experience de Recherche d'Objets Sombres) has been divided in two phases. 
The first one (1990-1995) was dedicated to a low mass macho
 search (no candidate were found in the mass range $10^{-7}$ to $10^{-3}$ $M_{\odot}$), 
and a medium sensitivity search to higher mass objects
 ($10^{-4}$ to 1 $M_{\odot}$). Results can be found in Renault et al (1997) and references therein.
The second phase started in Jun 1996 with the aim of improving the sensitivity by at least an order 
of magnitude for higher mass objects.
 It was conducted until Feb 2003, using a 1 meter dedicated telescope, the MARLY, situated at La Silla 
(Chile) and two wide field 32 million pixel CCD mosaics covering 1 deg$^2$. 
The total number of stars followed was 55 million (including 6 in the SMC), 
compared to 4 million LMC stars in the
 EROS-1 phase. The monitored solid angle was 88 (10) deg$^2$ in the LMC (SMC). 
(For comparaison, the MACHO collaboration monitored 11 million stars
 over 15 deg$^2$, mostly in the central region of the LMC~\cite{alcock2000a}). 
We decided to restrict our analysis to the 33 million brightest stars in our sample
 (29 in the LMC, 4 in the SMC). In order to analyse a homogeneous sample, 
all Magellanic Clouds images were photometered anew, using better and slightly
 deeper template images than in our previous reports. The typical sampling of light 
curves is about twice per week in each of two passbands. 

\section{Analysis and results}

The analysis conducted aimed at Einstein timescales shorter than two years. 
In this case, the microlensing light curves display a visible baseline flux.
After having applied a series of cuts to select single excursion from the baseline 
with a minimum signal-to-noise ratio, we then
 selected candidates by comparing the measurements with the best-fit point-source 
point-lens microlensing light curve; the last set of cuts deals with physical backgrounds,
 mainly "blue bumper" stars and supernovae. The former are easy to reject because of their 
limited and chromatic variations (larger in redder passbands), but the latter, which occur 
in galaxies behind the Magellanic Clouds (between 500 and 1000 Mpc), represent the dominant 
population in our sample of microlensing candidates at the end of the analysis : we found 28 
such objects, a rate comparable to that found by MACHO~\cite{alcock2000a}, preferentially in 
outer regions of the LMC. Supernovae are rejected by performing a fit of a function that 
includes an asymmetry parameter S, and reduces to the usual symmetric microlensing light curve 
when S=0, see Tisserand (2004). The function is obtained by replacing in the usual microlensing 
formula the Einstein timescale $t_E$ with $t_E[1+S\;arctan((t-t_0)/t_E)]$.
 We reject all candidates that have a large asymmetry parameter, $|S|>0.3$, or that are 
situated in the close neighbourhood of a visible galaxy
 (the probability of a chance alignment is very low). 
The tuning of each cut and the calculation of the microlensing detection efficiency are done 
with simulated simple microlensing light curves, 
as described in Palanque-Delabrouille et al. (1998). Combining our observation duration 
(6.7 years), the number of stars analysed and our average detection efficiency,  the sensitivity
 of this analysis is better by a factor 4 compared to the last search published by EROS-2 in the 
LMC~\cite{lasserre2000} and by a factor 2 with the one published by MACHO~\cite{alcock2000a}.

\subsection{Follow-up of the published microlensing candidates}

\begin{figure}
\begin{center}
\psfig{figure=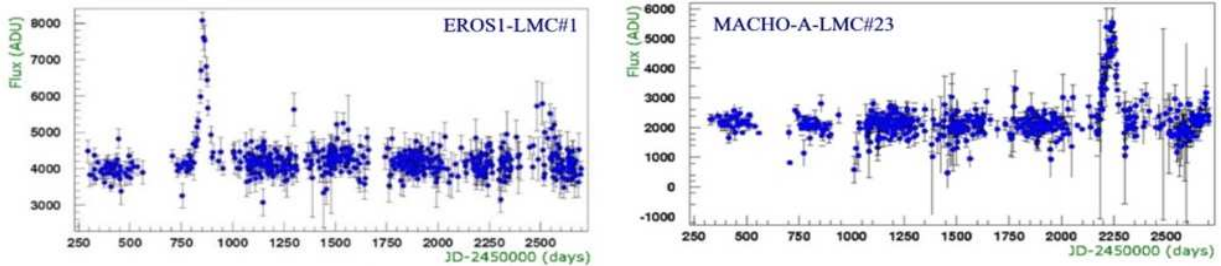,height=1.4in,width=6.4in}
\caption{EROS2 light curves (Jun 1996 - Feb 2003) of candidate EROS1-LMC\#1 (left) and 
MACHO-A-LMC\#23 (right) in the visible passband. The flux is in arbitrary units;
 the time is in days, with the origin at JD=2450000. Both bumps are new variations which 
appear 6.3 and 6.8 years after that originally seen by the EROS and MACHO group,
 before the startup of EROS-2. They are well fit by a microlensing light curve.}
\label{lc_variation}
\end{center}
\end{figure}

The first result came from the follow-up of the published microlensing candidates. 
The final EROS-2 data provides
an additional 6.6, 3.7 and 5.1 years time base compared to the latest publications 
by EROS1, Aubourg et al. (1993), EROS2, Milsztajn \& Lasserre (2001) and MACHO,
 Alcock et al. (2000a), useful to check for the stability of the baseline.\\
There were 5 surviving microlensing candidates from EROS towards the LMC, one from 
EROS-1 (number 1) and four from EROS-2 (numbered 3, 5, 6 and 7).
 Two of them (number 1 and 3) showed new variation incompatible with a microlensing 
effect. Candidate 1 displayed a new variation in 1998 (see figure \ref{lc_variation}),
 6.3 years after the first one, of similar amplitude (a factor two) and timescale (28 days). 
This second variation is well fit by a microlensing light curve, but because variations
 are separated in time by more than 80 Einstein timescales, the probability that these two 
bumps correspond to the microlensing of a double source star is lower than half a percent.
 We also attempted to follow all 13 MACHO candidates selected by their analysis A 
(stricter cuts). We were able to identify unambiguously nine of them and one star,
 MACHO-A-LMC\#23, displays a new variation in Dec 2001, very similar to that observed by 
the MACHO group in Feb 1995 (variation of 1 magnitude and timescale of 40 days)
 (see figure \ref{lc_variation}).

The second result came from the improved photometry and the refined cut use to reject 
supernovae. The light curves of candidates 5, 6 and 7 show an asymmetry between the
rise and fall times; they correspond to supernovae. Note that candidate 5 is identical 
to MACHO candidate 26, which was rejected by Alcock et al. (2000a)
 for the same reason.

The conclusion is that none of the former EROS microlensing candidates published towards
the LMC are still considered valid.

\subsection{LMC results}

The present analysis of the LMC data has selected four new microlensing candidates. 
One, EROS2-LMC\#8, is a high signal-to-noise microlensing event which shows a chromatic
 variation of a factor 12 (25) in the red (visible) band. Its position in the 
colour-magnitude diagram is abnormal, almost 1 mag redder than comparable LMC objects in 
the main sequence. When a possible blending is taken into account however, the 
microlensing fit becomes excellent, with agreement between the timescales in the two 
passbands. The baseline flux corresponding
 to the magnified object appears to be within the dimmer part of the observed LMC main 
sequence; this is compatible with the source being an LMC star. In contrast, the non-ampified 
component is now redder, and cannot correspond to an LMC star. A good solution is that 
the unmagnified flux corresponds to the lens, a 13.5 mag M dwarf (absolute mag) at a distance 
of 300 pc. The three other candidates, EROS2-LMC\#9 to 11, have lower signal-to-noise. Their 
source stars have baseline flux at about 21st mag and the timescales range from 35 to 55 days. 

\begin{figure}
\begin{center}
\psfig{figure=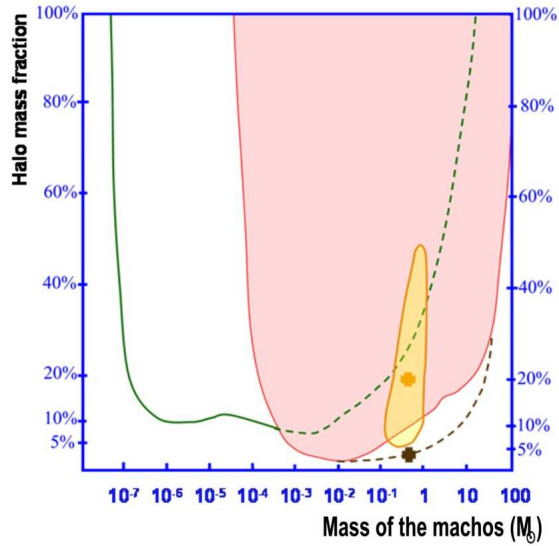,height=3.0in}
\caption{The preliminary 95\% C.L. upper limit on the macho content of the halo obtained from 
the final EROS2 LMC data only (red curve between $10^{-4}$ and 100 $M_{\odot}$).
 The lower cross is the value of the optical depth towards the LMC corresponding to the three 
candidates. The dashed line that goes through this cross shows what the limit would be
 in the absence of microlensing candidates. This limit is compared to that presented in Lasserre 
et al. (2000) (green curve between $10^{-7}$ and 10 $M_{\odot}$).
 The smaller orange-shaded contour between 0.1 and 1 $M_{\odot}$ is the signal presented by Alcock 
et al. (2000a), with the lighter cross indicating their preferred solution. It must be recalled
 that the fields monitored by MACHO and EROS2 are not identical.}
\label{diag_limit}
\end{center}
\end{figure}

As the event EROS2-LMC\#8 is most likely due to a galactic disk lens, the optical depth has 
been calculated with the remaining 3 candidates, if interpreted as machos. The value obtained
 is 1.5 $10^{-8}$, i.e. 3\% of the standard spherical halo. A preliminary version of a 95\% C.L. 
upper limit on the macho content of the halo is presented in Fig. \ref{diag_limit}.
 It uses only the EROS-2 LMC data set; combination with both the EROS1 and the EROS2 SMC results 
will be done later.  The limit is better than 12\% of the halo for macho masses
 between $2.10^{-4}$ and 1 M$_\odot$.

\section{Conclusion}

The first account of the analysis of the full EROS-2 data set towards the LMC was presented. 
This data set has presently the largest sensitivity to halo machos. The small number of
microlensing candidates is lower than expected from the results of the MACHO group.  From 
this, one derives strict limits on the abundance of machos. A few microlensing candidates
have been shown to vary again, many years later; they should be removed from the candidates census.

\end{document}